\documentclass[Physsubmission, Phys]{SciPost}

\hypersetup{
    colorlinks,
    linkcolor={red!50!black},
    citecolor={blue!50!black},
    urlcolor={blue!80!black}
}

\usepackage[bitstream-charter]{mathdesign}
\usepackage{geometry}
\DeclareSymbolFont{usualmathcal}{OMS}{cmsy}{m}{n}
\DeclareSymbolFontAlphabet{\mathcal}{usualmathcal}

\begin{document}
\begin{center}{\Large \textbf{
Intermittency analysis in relativistic heavy-ion collisions\\
}}\end{center}

\begin{center}
Zhiming Li\textsuperscript{$\star$} and Jin Wu
\end{center}

\begin{center}
Key Laboratory of Quark and Lepton Physics (MOE) and Institute of Particle Physics, Central China Normal University, Wuhan 430079, China\\

* lizm@mail.ccnu.edu.cn
\end{center}

\begin{center}
\today
\end{center}


\definecolor{palegray}{gray}{0.95}
\begin{center}
\colorbox{palegray}{
  \begin{tabular}{rr}
  \begin{minipage}{0.1\textwidth}
    \includegraphics[width=30mm]{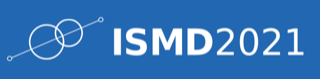}
  \end{minipage}
  &
  \begin{minipage}{0.75\textwidth}
    \begin{center}
    {\it 50th International Symposium on Multiparticle Dynamics}\\ {\it (ISMD2021)}\\
    {\it 12-16 July 2021} \\
    \doi{10.21468/SciPostPhysProc.?}\\
    \end{center}
  \end{minipage}
\end{tabular}
}
\end{center}

\section*{Abstract}
{\bf
Local density fluctuations near the QCD critical point can be probed by an intermittency analysis of power-law behavior on scaled factorial moments in relativistic heavy-ion collisions. We study the second-order scaled factorial moment in Au + Au collisions at $\sqrt{s_\mathrm{NN}}$ = 7.7-200 GeV from the UrQMD model. Since the background subtraction and efficiency correction are two important aspects in this measurement, we propose a cumulative variable method to remove background contribution and a cell-by-cell method for efficiency correction in the intermittency analysis.
}

\vspace{10pt}
\noindent\rule{\textwidth}{1pt}
\tableofcontents\thispagestyle{fancy}
\noindent\rule{\textwidth}{1pt}
\vspace{10pt}

\section{Introduction}
\label{sec:intro}
One of the main goals of relativistic heavy-ion collisions is to explore the QCD phase diagram and phase boundary~\cite{QCDReport}. An important feature of the QCD phase diagram is the critical point which is supposed to be a unique property of strongly interacting matter~\cite{CEP2}. In analogue to the critical opalescence observed in conventional matter near critical point, the related fractal geometry of QCD matter gives rise to local density fluctuations that obey intermittency behavior~\cite{AntoniouPRL}. In the NA49 experiment~\cite{NA49EPJC}, a power-law behavior has been observed in Si + Si collisions at 158A GeV~\cite{NA49EPJC}. In a preliminary analysis of the STAR experiment, the scaling exponent extracted from intermittency index shows a minimum in central Au + Au collisions around $\sqrt{s_\mathrm{NN}}$ = 20-30 GeV~\cite{STARintermittency}. Meanwhile, models which involve self-similar property~\cite{CMCPLB} and hadronic potentials~\cite{UrQMDLi} show that the intermittency behavior could be observed in Au + Au collisions at RHIC energies.
 
In the measurement of intermittency, it is crucial to estimate and remove background effects which significantly influence the particle multiplicity spectra in finite space of high-energy collisions. In the mean time, the measured  scaled factorial moments of multiplicity distributions in momentum space are different from the original ones due to experimental detection efficiency~\cite{STARPRCMoment}. In this work, we will study how to remove the trivial background effects by a cumulative variable method, and how to recover the scaled factorial moments from the experimentally measured ones by applying a proper efficiency correction method.

\section{The Method}
In heavy-ion collisions, the power-law fluctuations near the QCD CP can be detectable in momentum space through the measurement of the scaled factorial moment (SFM) defined as:\\
\begin{equation}
F_{q}(M)=\frac{\langle\frac{1}{M^{D}}\sum_{i=1}^{M^{D}}n_{i}(n_{i}-1)\cdots(n_{i}-q+1)\rangle}{\langle\frac{1}{M^{D}}\sum_{i=1}^{M^{D}}n_{i}\rangle^{q}},
 \label{Eq:FM}
\end{equation}

\noindent where $M$ is the number of equally sized cells in one of D-dimensional space, $n_{i}$ is the measured multiplicity in the $i$-th cell, and $q$ is the order of moment.

A power-law (scaling) behavior of $F_{q}(M)$ on the number of cells $M^{D}$ when $M$ is large enough, is referred to as intermittency behavior:

\begin{equation}
F_{q}(M)\sim (M^{D})^{\phi_{q}}, M\rightarrow\infty.
 \label{Eq:PowerLaw}
\end{equation}

\noindent Here the exponent $\phi_q$ is the intermittency index.

In this work, we investigate the second-order SFM, $F_{2}(M)$, of proton density in transverse momentum space ($p_{x}-p_{y}$) in Au $+$ Au collisions at $\sqrt{s_\mathrm{NN}}$ = 7.7-200 GeV from the UrQMD model.

\section{Results and Discussions}
\label{sec:another}
\subsection{Energy Dependence of Scaled Factorial Moments }
\begin{figure}[!htp]
     \centering
     \includegraphics[scale=0.55]{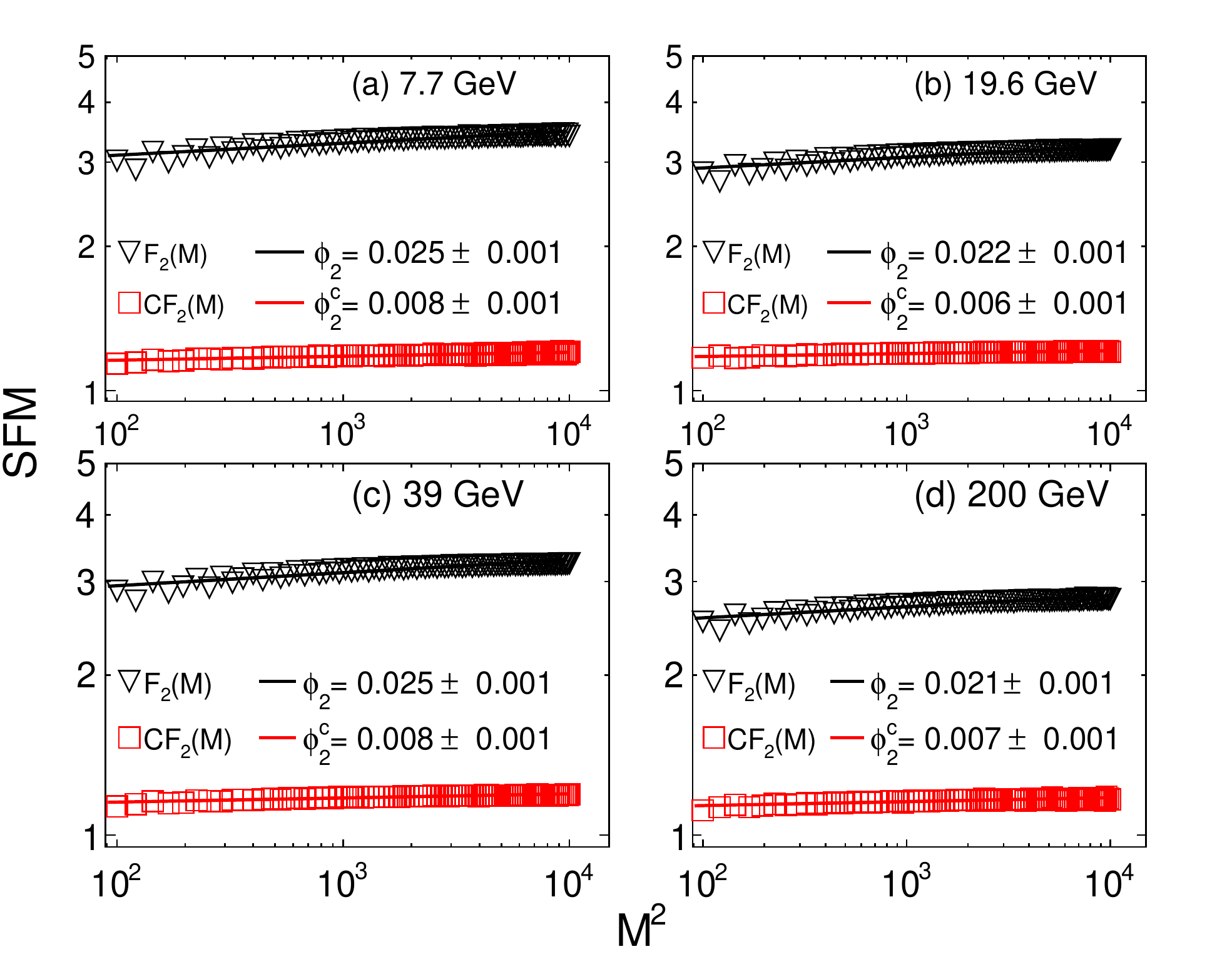}
      \caption{$F_{2}(M)$ as a function of $M^{2}$ in Au + Au collision from the UrQMD model. The red symbols represent the SFMs calculated by the cumulative variable method.}
     \label{Fig:F2E}
\end{figure}

In figure 1, we show the $F_{2}(M)$ as a function of $M^{2}$ in Au + Au collision at $\sqrt{s_\mathrm{NN}}$ = 7.7, 19.6, 39 and 200 GeV from the UrQMD model.  More details have been presented in Ref~\cite{RefEfficiency}. It is observed that $F_{2}(M)$ increases slowly at small $M^{2}$ region and becomes flat when $M^{2}$ is large. The intermittency indices $\phi_{2}$ which obtained from the fitting of Eq.~\eqref{Eq:PowerLaw}, are small at all energies. Therefore, We find that $\phi_{2}$ from the directly calculated SFMs are small but nonzero. However, this result is hard to be understood since $\phi_{2}$ is expected to be zero because the UrQMD model used here does not include any critical density fluctuations. Therefore, there must exist unexpected background which contributes to the value of $\phi_{2}$. In the next section, we will study how to remove the background effects by a proposed method.

\subsection{Background Subtraction}
The cumulative variable $X(x)$ is calculated from the single-particle density distribution $\rho(x)$ and is defined as ~\cite{BialasCumulative}:

\begin{equation}
  \large X(x)=\frac{\int_{x_{min}}^{x} \rho(x)dx}{\int_{x_{min}}^{x_{max}}\rho(x)dx}.
 \label{Eq:cvariable}
\end{equation}

\noindent Where $x$ represents the original variable and $\rho(x)$ is the density function of $x$. $X(x)$ is the new variable after a cumulative transformation.  

\begin{figure}[htp]
     \centering
     \includegraphics[scale=0.55]{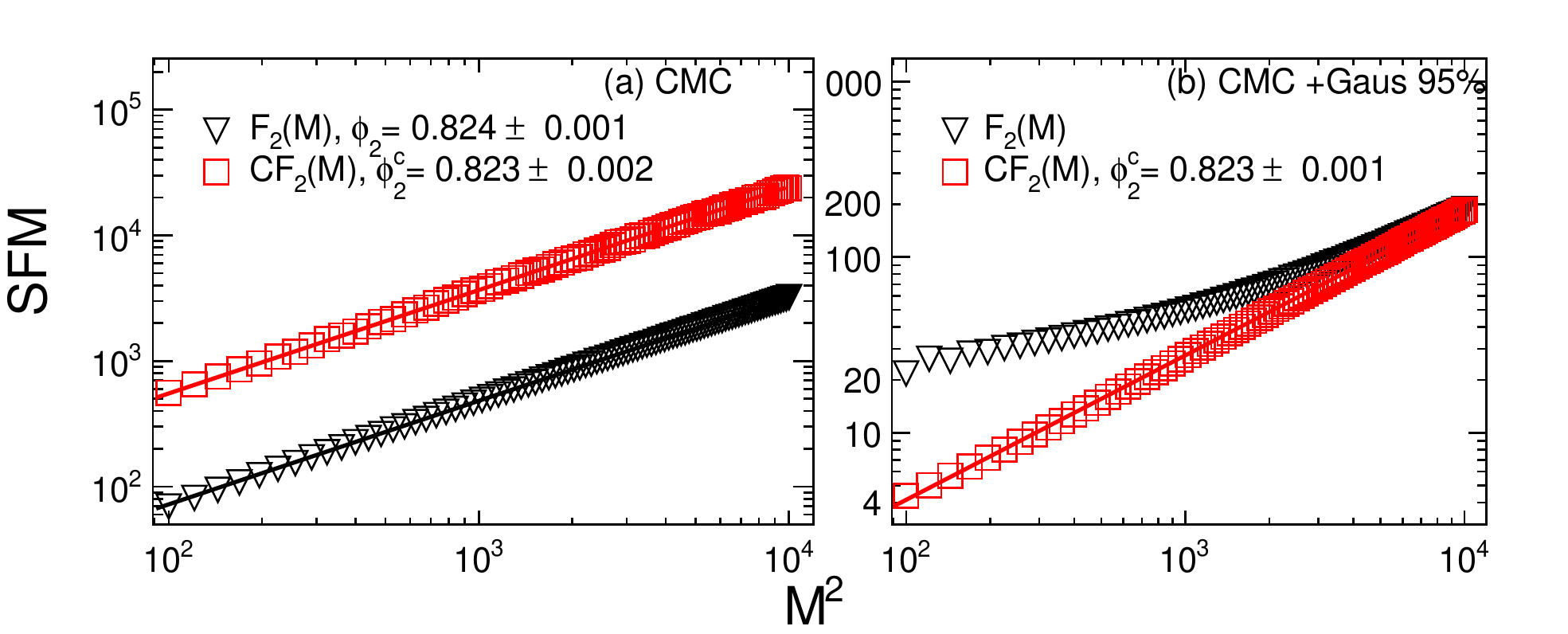}
     \label{Fig:GasandCMC}
     \caption{(a) $F_{2}(M)$ and $CF_{2}(M)$ as a function of $M^{2}$ in pure CMC events. (b) The result in CMC events contaminated with Gaussian background fluctuations.}
\end{figure}

After using the cumulative variable, the $F_{2}(M)$ will transfer to $CF_{2}(M)$. To check validity of the cumulative variable method in the intermittency analysis, we use a critical Monte Carlo (CMC) model~\cite{AntoniouPRL,CMCPLB} to simulate critical events. In figure 2(a), it is clearly observed that both the $F_{2}(M)$ and $CF_{2}(M)$ obey a good power-law behavior with $M^{2}$. Then, we let CMC event samples contaminated with background events from Gaussian statistical distributions. In figure 2(b), it is found that $CF_{2}(M)$ still obeys a power-law behavior with $M^{2}$ but $F_{2}(M)$ fails. 

In figure 1, we also show the SFMs calculated by the cumulative method (red symbols) and the $\phi_{2}^{c}$ extracted from the fit of $CF_{2}(M)$. It is observed that $\phi_{2}^{c}$ is much smaller than $\phi_{2}$. These results verify that the background of noncritical effect can be efficiently removed by the proposed cumulative variable method.

\subsection{Efficiency Correction}
To recover the SFM from efficiency effects, we assume the response of the detector follows a binomial probability distribution function. Then the directly measured SFM can be corrected by~\cite{RefEfficiency} 
 
\begin{eqnarray}
  F_{q}^{corrected}(M)=\frac{\langle\frac{1}{M^{2}}\sum_{i=1}^{M^{2}}\frac{n_{i}(n_{i}-1)\cdots(n_{i}-q+1)}{\bar{\epsilon_{i}}^{q}}\rangle}{\langle\frac{1}{M^{2}}\sum_{i=1}^{M^{2}}\frac{n_{i}}{\bar{\epsilon_{i}}}\rangle^{q}}.
 \label{Eq:FMcorrection}
 \end{eqnarray}

\noindent where $\bar{\epsilon_{i}}$ is the event average of the mean efficiency for the particles located in the $i$th cell. 

\begin{figure}[!htp]
      \centering
     \includegraphics[scale=0.55]{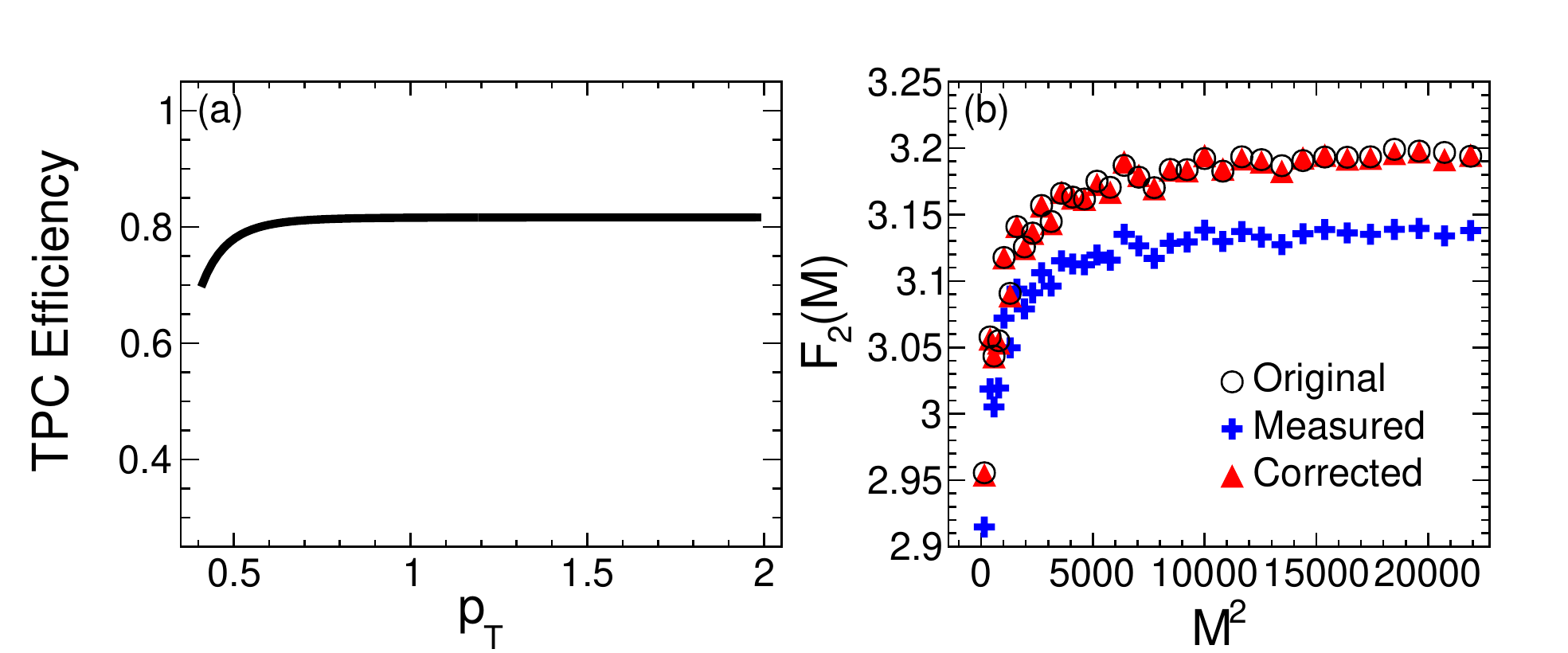}
      \caption{(a) TPC efficiency as a function of $p_{T}$ for protons in Au $+$ Au collisions at the STAR experiment~\cite{STARPRCMoment}. (b) The original , measured and efficiency corrected $F_{2}(M)$ as a function of $M^{2}$ at $\sqrt{s_\mathrm{NN}}$ = 19.6 GeV from the UrQMD model.}
 \label{Fig:F2TPC}
 \end{figure} 

Figure 3 (a) shows the tracking efficiency as a function of $p_{T}$ for protons in TPC detector at STAR~\cite{STARPRCMoment}. After randomly discarding particles according to this $p_{T}$-dependent efficiency, we could get the measured event sample. In Figure 3 (b), the measured $F_{2}(M)$ are obviously smaller than the original ones. But the efficiency corrected $F_{2}(M)$, which calculated based on Eq.~\eqref{Eq:FMcorrection}, are almost overlap with the true original ones. Therefore, the cell-by-cell method works well for the efficiency correction of SFM. 

\section{Summary}
In this work, we study the intermittency in Au $+$ Au collisions at $\sqrt{s_\mathrm{NN}}=7.7 - 200$ GeV by using the UrQMD model. The second-order intermittency indices obtained from the directly measured SFMs, are found to be small but nonzero at all energies. The cumulative variable method is proposed to effectively reduce the distortion of a Gaussian background and the non-critical background effects in the UrQMD model. As for the efficiency correction on SFM, we propose a cell-by-cell method and it is found to be valid by introducing the STAR detector tracking efficiency to the UrQMD event sample. Our results provide a noncritical baseline and could give a guidance to background subtraction and efficiency correction for the intermittency analysis in heavy-ion experiments.  

\section*{Acknowledgements}
This work is supported by the Innovation Fund of the Key Laboratory of Quark and Leptonic Physics (Grant No. QLPL2020P01) and the Ministry of Science and Technology (MoST) under grant No. 2016YFE0104800.
\vspace{-0.5cm}
\bibliography{SciPost_Example_BiBTeX_File.bib}

\begin{thebibliography}{10}
\providecommand{\url}[1]{\texttt{#1}}
\providecommand{\urlprefix}{URL }
\expandafter\ifx\csname urlstyle\endcsname\relax
  \providecommand{\doi}[1]{doi:\discretionary{}{}{}#1}\else
  \providecommand{\doi}{doi:\discretionary{}{}{}\begingroup
  \urlstyle{rm}\Url}\fi
\providecommand{\eprint}[2][]{\url{#2}}

\bibitem{QCDReport}
A.~Bzdak \emph{et~al.},
\newblock \emph{{Mapping the Phases of Quantum Chromodynamics with Beam Energy
  Scan}},
\newblock Phys. Rept. \textbf{853}, 1 (2020),
\newblock \doi{10.1016/j.physrep.2020.01.005}.

\bibitem{CEP2}
Y.~Hatta and M.~A. Stephanov,
\newblock \emph{{Proton number fluctuation as a signal of the QCD critical
  endpoint}},
\newblock Phys. Rev. Lett. \textbf{91}, 102003 (2003),
\newblock \doi{10.1103/PhysRevLett.91.102003}.

\bibitem{AntoniouPRL}
N.~G. Antoniou \emph{et~al.},
\newblock \emph{{Critical opalescence in baryonic QCD matter}},
\newblock Phys. Rev. Lett. \textbf{97}, 032002 (2006),
\newblock \doi{10.1103/PhysRevLett.97.032002}.

\bibitem{NA49EPJC}
T.~Anticic~{\it et al.} (NA49~Collaboration),
\newblock \emph{{Critical fluctuations of the proton density in A+A collisions
  at 158$A$ GeV}},
\newblock Eur. Phys. J. C \textbf{75}(12), 587 (2015),
\newblock \doi{10.1140/epjc/s10052-015-3738-5}.

\bibitem{STARintermittency}
J.~Wu~(for~the STAR~Collaboration),
\newblock \emph{{Measurement of Intermittency for Charged Particles in Au + Au
  Collisions at $\sqrt{s_\mathrm{NN}}$ = 7.7-200 GeV from STAR}},
\newblock presentation at ISMD  (2021).

\bibitem{CMCPLB}
J.~Wu \emph{et~al.},
\newblock \emph{{Probing QCD critical fluctuations from intermittency analysis
  in relativistic heavy-ion collisions}},
\newblock Phys. Lett. B \textbf{801}, 135186 (2020),
\newblock \doi{10.1016/j.physletb.2019.135186}.

\bibitem{UrQMDLi}
P.~Li \emph{et~al.},
\newblock \emph{{Proton correlations and apparent intermittency in the UrQMD
  model with hadronic potentials}},
\newblock Phys. Lett. B \textbf{818}, 136393 (2021),
\newblock \doi{10.1016/j.physletb.2021.136393}.

\bibitem{STARPRCMoment}
M.~Abdallah~{\it et al.} (STAR~Collaboration),
\newblock \emph{{Cumulants and correlation functions of net-proton, proton, and
  antiproton multiplicity distributions in Au+Au collisions at energies
  available at the BNL Relativistic Heavy Ion Collider}},
\newblock Phys. Rev. C \textbf{104}(2), 024902 (2021),
\newblock \doi{10.1103/PhysRevC.104.024902}.

\bibitem{RefEfficiency}
J.~Wu \emph{et~al.},
\newblock \emph{{Intermittency analysis of proton numbers in heavy-ion
  collisions at energies available at the BNL Relativistic Heavy Ion
  Collider}},
\newblock Phys. Rev. C \textbf{104}(3), 034902 (2021),
\newblock \doi{10.1103/PhysRevC.104.034902}.

\bibitem{BialasCumulative}
M.~G. A.~Bialas,
\newblock \emph{{A new variable to study intermittency}},
\newblock Physics Letters B \textbf{252}, 483 (1990),
\newblock \doi{https://doi.org/10.1016/0370-2693(90)90575-Q}.

\end{thebibliography}

\nolinenumbers

\end{document}